\documentclass[sigconf,natbib=true]{acmart} 
\usepackage{caption}
\begin{document}

\title{Towards Sustainable  Personalized On-Device Human Activity Recognition with TinyML and Cloud-Enabled Auto Deployment}

\author{Bidyut Saha}
\email{bidyutsaha@kgpian.iitkgp.ac.in}
\affiliation{%
  \institution{Indian Institute of Technology Kharagpur}
  \country{India}
}

\author{Riya Samanta}
\email{riya.samanta@iitkgp.ac.in}
\affiliation{%
  \institution{Indian Institute of Technology Kharagpur}
  \country{India}
}

\author{Soumya K. Ghosh}
\email{skg@cse.iitkgp.ac.in}
\affiliation{%
  \institution{Indian Institute of Technology Kharagpur}
  \country{India}
}

\author{Ram Babu Roy}
\email{rambabu@see.iitkgp.ac.in}
\affiliation{%
  \institution{Indian Institute of Technology Kharagpur}
  \country{India}
}

\begin{abstract}
  Human activity recognition (HAR) holds immense potential for transforming health and fitness monitoring, yet challenges persist in achieving personalized outcomes and sustainability for on-device continuous inferences. This work introduces a wrist-worn smart band designed to address these challenges through a novel combination of on-device TinyML-driven computing and cloud-enabled auto-deployment. Leveraging inertial measurement unit (IMU) sensors and a customized 1D Convolutional Neural Network (CNN) for personalized HAR, users can tailor activity classes to their unique movement styles with minimal calibration. By utilising TinyML for local computations, the smart band reduces the necessity for constant data transmission and radio communication, which in turn lowers power consumption and reduces carbon footprint. This method also enhances the privacy and security of user data by limiting its transmission. Through transfer learning and fine-tuning on user-specific data, the system achieves a 37\% increase in accuracy over generalized models in personalized settings. Evaluation using three benchmark datasets — WISDM, PAMAP2, and the BandX — demonstrates its effectiveness across various activity domains. Additionally, this work presents a cloud-supported framework for the automatic deployment of TinyML models to remote wearables, enabling seamless customization and on-device inference, even with limited target data. By combining personalized HAR with sustainable strategies for on-device continuous inferences, this system represents a promising step towards fostering healthier and more sustainable societies worldwide.
\end{abstract}



\maketitle

\footnote*{This poster has been derived from the paper "Bidyut Saha, et al. TinyML-Driven On-Device Personalized Human Activity Recognition and Auto-Deployment to Smart Bands. In Proceedings of The Third International Conference on Artificial Intelligence and Machine Learning Systems (AIML-Systems 2023). Bengaluru, India. It has received the Best Paper Award at the AI-ML Systems Conference.}

\section{Introduction}
Human activity Recognition (HAR) involves employing learning-based algorithms to automatically identify activities by analyzing sensor data from wearables and environmental sensors. HAR holds wide-ranging applications, particularly with the rising popularity of wearable fitness trackers. However, current HAR approaches often rely on computational offloading, leading to drawbacks such as slow response times, high power consumption, and dependency on network availability.

To address these challenges, on-device computation has emerged as a solution. However, on-device computation among wearables faces constraints such as limited computational resources and memory. Additionally, traditional HAR models focus on generalization rather than personalization, leading to reduced performance for new users and or for tailored behavioural patterns.

This study addresses these issues by evaluating a personalized HAR model using benchmark datasets and transfer learning techniques. We propose a lightweight 1D Convolutional Neural Network (CNN) architecture for HAR deployable on resource-constrained Microcontroller Units (MCUs) using TinyML. Additionally, we develop an end-to-end framework for personalized on-device HAR with minimal user intervention, allowing for over-the-air auto-deployment onto wrist-worn activity trackers.

To facilitate user interaction and data annotation, we have developed a web application hosted on the cloud. Users can create accounts, link their devices, and add personalized fitness classes. The accuracy and performance of the model are evaluated using benchmark datasets.

In terms of sustainability, our approach reduces the reliance on external computational resources and minimizes data transmission, leading to lower energy consumption and reduced environmental impact related to carbon emissions. By enabling personalized on-device inference, we promote long-term viability and reduce dependency on centralized infrastructure.

The paper is structured as follows: Section \ref{relatedwork} reviews related literature, Section \ref{method} gives the implementation details of the proposed HAR approach and its deployment, Section \ref{exp} discusses experimental analysis, and Section \ref{conc} concludes the study.

\section{Related Work}
\label{relatedwork}

\subsection{On-device HAR}
In recent literature, various studies have explored the implementation of on-device Human Activity Recognition (HAR) systems across different platforms. Zebin et al. \cite{zebin2019design} developed a HAR system utilizing a CNN deployed on a mid-range smartphone, leveraging its accelerometer and gyroscope sensors to detect activities like walking upstairs, downstairs, sedentary behaviour, and sleep patterns. This model, trained initially on a standalone computer using TensorFlow, was optimized for smartphone deployment through the TensorFlow Lite library \cite{tensorflowlite}.

Similarly, Mairittha et al. \cite{mairittha2019device} employed a Long Short-Term Memory (LSTM) model for real-time activity detection using the locomotion sensors inherent in smartphones. Coelho et al. \cite{coelho2021lightweight} proposed a lightweight HAR framework powered by a Microcontroller Unit (MCU), employing a two-stage classification approach. Their system distinguished between static and dynamic actions using hand-crafted features and a Decision Tree at the first level, followed by employing either another Decision Tree or a CNN based on action complexity.

Alessandrini et al. \cite{alessandrini2021recurrent} constructed a classifier using a Recurrent Neural Network (RNN) module with data from Photoplethysmography (PPG), accelerometer, and gyroscope sensors, which was deployed on an MCU development board. Daghero et al. \cite{daghero2022two} introduced a two-stage HAR system aiming to enhance energy efficiency by combining a small Decision Tree with a 1D CNN deployed on an MCU unit.

In a notable prototype, Saha and colleagues \cite{saha2023bandx} integrated a wrist-worn human activity tracker with an on-device MCU unit employing a 1D CNN to analyze embedded Inertial Measurement Unit (IMU) sensor data. Activity labels were transmitted to a cloud environment for activity history maintenance. However, it is noteworthy that while these HAR models provide valuable insights, they are generalized for user groups rather than personalized for individual users.

\subsection{Personalized HAR}

The challenge in utilizing wearable sensors for human activity recognition lies in the diminished performance of recognition models when new users with various physical or behavioural traits are introduced. Existing activity recognition systems typically employ generalized methods to categorize human actions, which struggle to accurately classify actions of new users due to the variability in human behaviour and action execution \cite{khowaja2020caphar}. To address this variability, personalized models have emerged, showcasing the potential to enhance the accuracy of machine learning and deep learning algorithms.

Ferrari et al. \cite{ferrari2020personalization} introduced the concept of subject similarity, devising a technique to assign scores to subjects based on their resemblance to others. They personalized the model for each individual by prioritizing data from subjects exhibiting the highest degree of similarity during model training. Similarly, Khowaja et al. \cite{khowaja2020caphar} proposed individualized activity detection by creating distinct models for each user. They adopted a two-step process, identifying the most similar users from the dataset and utilizing the model trained by these users to personalize the model for new users.

Vu et al. \cite{vu2019understanding} suggested a classifier selection approach based on compatibility, training various classifiers on a group of subjects and selecting the most suitable one for a target subject based on performance metrics. Rokni et al. \cite{rokni2018personalized} developed a HAR model using Convolutional Neural Networks (CNNs), initially trained with data from a source domain. They transferred this pre-trained model to build personalized HAR models for new users in a target domain using transfer learning techniques, allowing for personalized model creation with minimal supervision.

However, while these studies contribute to the theoretical aspects of HAR model personalization, they lack insights into real-time implementation aspects. Hence, there is a crucial need to develop personalized and efficient on-device systems for HAR that optimize trade-offs between latency, energy consumption, implementation cost, real-time performance, data privacy, and accuracy, while considering users' unique activity features, movement patterns, habits, and preferences.

Our work aims to bridge this gap by providing end-to-end TinyML-based on-device implementation details, offering insights into the practical deployment of personalized HAR systems, particularly focusing on low-resource devices. By addressing the challenges of real-time implementation, our research endeavors to facilitate the development of personalized and effective on-device HAR systems, ultimately promoting healthier and more active lifestyles.

\begin{figure}[!t]
 \centering
  \includegraphics[width=0.9\linewidth]{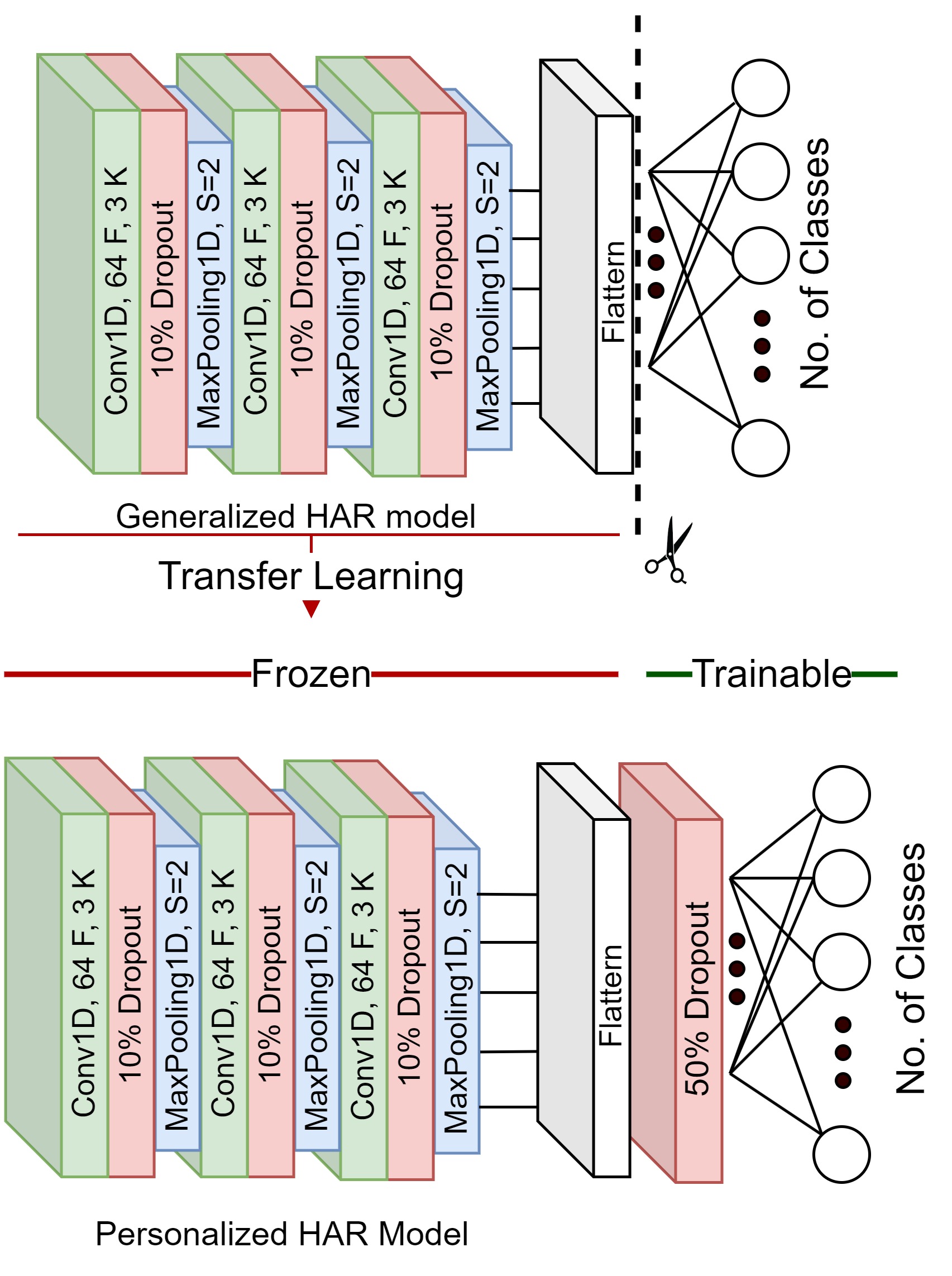} 
  \vspace{-0.1in}
  \caption{Architecture of the 1D CNN for personalized Human Activity Recognition (HAR). (Source: Bidyut Saha et al. "TinyML-Driven On-Device Personalized Human Activity Recognition and Auto-Deployment to Smart Bands", AIML-Systems 2023)}
  \label{neuralArchitecture}
  \vspace{-0.2in}
\end{figure}

\section{System Implementation}
\label{method}
The proposed prototype builds upon the foundational technology of the BandX \cite{saha2023bandx} device, evolving it into a software-upgraded version. The BandX device, initially designed for on-device computing, leveraged TinyML to implement a lightweight 1D CNN model on a resource-constrained MCU. This setup enabled the device to track human activities by analyzing deviations in data from an embedded IMU sensor. Worn on the wrist, BandX is capable of classifying seven types of human locomotion, which include walking, jogging, cycling, typing, writing, ascending stairs, and descending stairs. The core of the BandX utilized an ESP32 Vroom as its processing unit, boasting 320KB of SRAM, 4MB of Flash memory, and built-in WiFi and BLE connectivity. An accelerometer and gyroscopic sensor module were interfaced with it, alongside control buttons and an OLED display. A battery and charge controller module were integrated, and the entire setup was enclosed within a 3D-printed wristband enclosure.

This work introduces two novel software features to enhance the BandX system. Firstly, personalized HAR is implemented through transfer learning with fine-tuning techniques. Initially, a model is trained with HAR data from multiple users, serving as a pre-trained model for subsequent fine-tuning specific to the target user. This involves employing a 1D CNN model, the architecture of which is illustrated in Figure \ref{neuralArchitecture}. To personalize the model, only the last layer is fine-tuned using a small amount of target data.

\begin{figure}[!t]
 \centering
  \includegraphics[width=\linewidth]{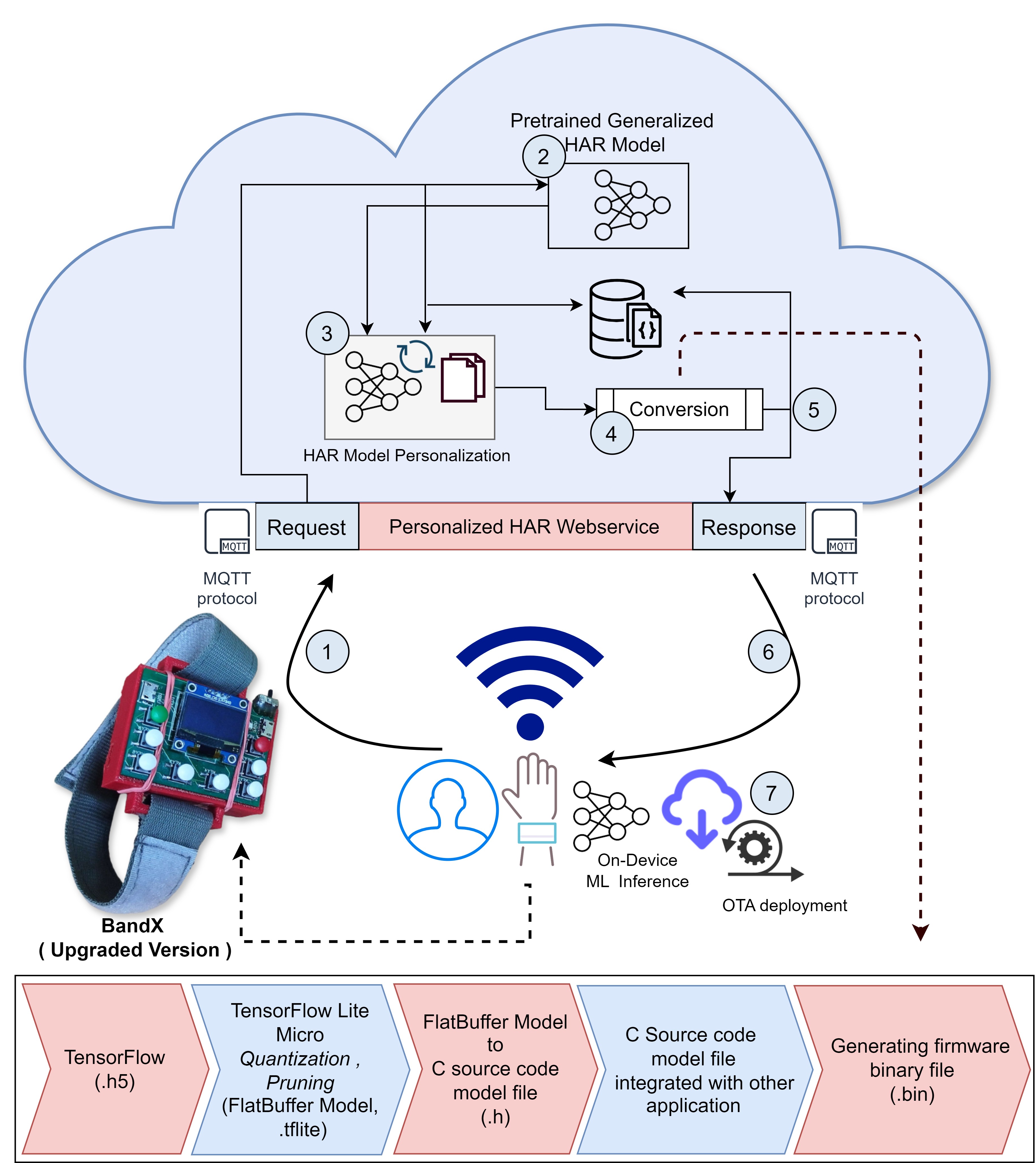} 
  \vspace{-0.1in}
  \caption{Cloud-assisted auto-deployment workflow framework for personalized HAR. (Source: Bidyut Saha et al. "TinyML-Driven On-Device Personalized Human Activity Recognition and Auto-Deployment to Smart Bands", AIML-Systems 2023)}
  \label{framework}
  \vspace{-0.1in}
\end{figure}

\begin{figure}[h]
 \centering
  \includegraphics[width=\linewidth]{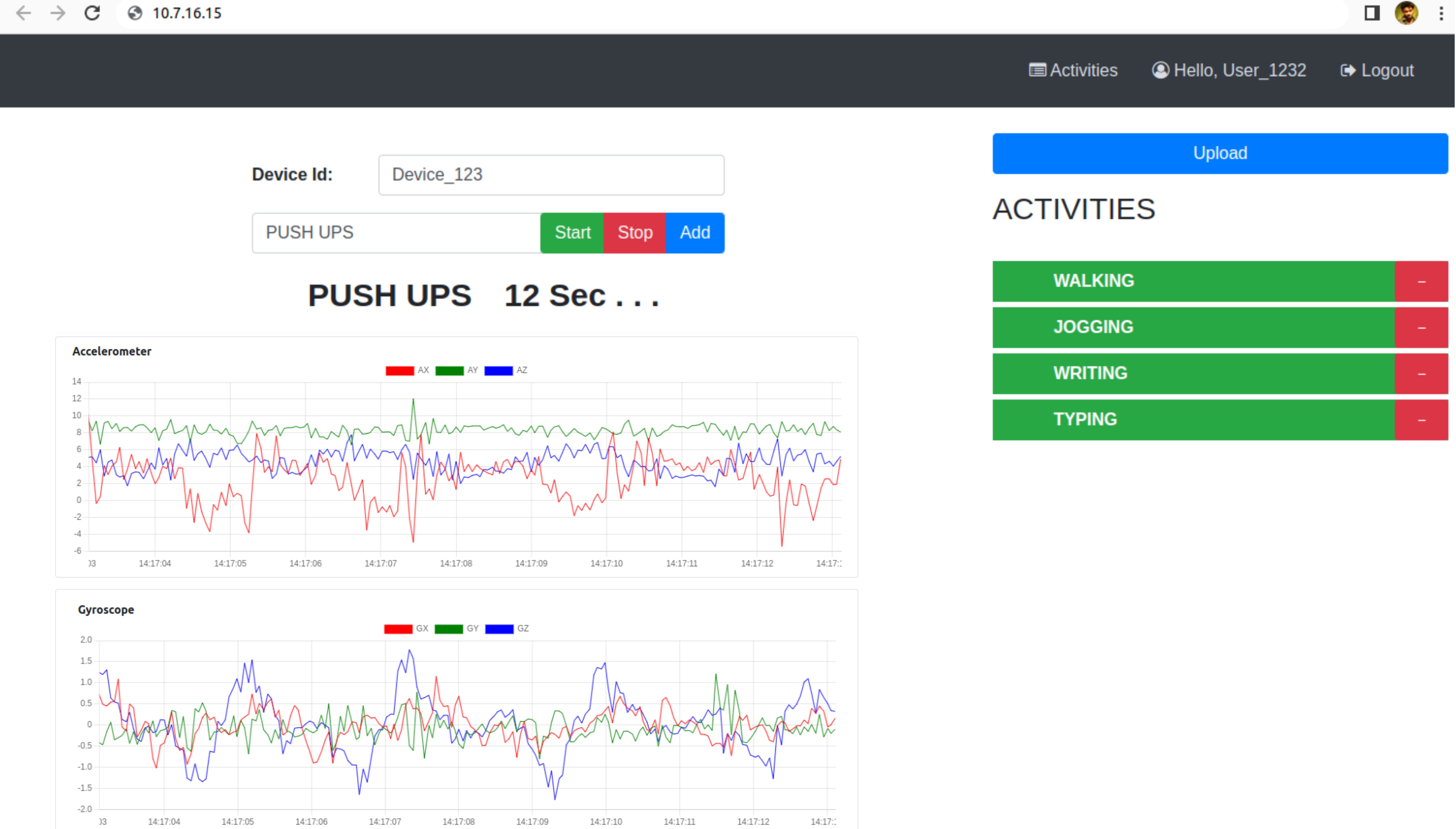} 
  \vspace{-0.1in}
  \caption{Dashboard for uploading user data for personalized HAR.(Source: Bidyut Saha et al. "TinyML-Driven On-Device Personalized Human Activity Recognition and Auto-Deployment to Smart Bands", AIML-Systems 2023)}
  \label{dashboard}
  \vspace{-0.1in}
\end{figure}

The second enhancement involves the automatic deployment of the fine-tuned model to the target device, as detailed in Figure \ref{framework}. Initially, users upload a small amount of data for various classes of interest via a web request, labelling it through the web application depicted in Figure \ref{dashboard}. This data is then used to fine-tune a pre-trained model in the cloud, creating a personalized model for each user.

Subsequently, the retrained model undergoes conversion and optimization stages to make it deployable on resource-constrained MCUs. TensorFlow \cite{abadi2016tensorflow,david2021tensorflow}, a machine learning framework, is utilized for training the models. The fine-tuned model is temporarily stored in the cloud environment before being converted into TensorFlow Lite format \cite{tensorflowlite}. This conversion incorporates techniques such as pruning and integer quantization to ensure compatibility with resource constraints.

For deployment on MCUs and similar devices, TensorFlow Lite Micro \cite{tensorflowlitemicro}, tailored for running machine learning models in resource-constrained environments, is employed. The bundled TensorFlow Lite model, along with the code, is prepared to update the firmware of the BandX device.

Upon receiving the response from the previous request, the upgraded BandX device initiates the firmware update process, enabling over-the-air deployment. This seamless integration ensures the efficient incorporation of personalized HAR capabilities.

\section{Experimental Analysis}
\label{exp}
\subsection{Datasets}\label{data}
\noindent\textbf{Description:} Three datasets are employed in this study. WISDM \cite{weiss2019smartphone} and PAMAP2 \cite{reiss2012introducing} are publicly available datasets. The third dataset is obtained through crowdsourcing using the BandX device \cite{saha2023bandx}. Details about these datasets are summarized in Table \ref{dataset}.

\begin{table}[htpb]
\centering
\caption{Datasets used in our study to evaluate the performance of proposed personalized HAR system.}
\label{dataset}
\vspace{-0.1in}
\setlength{\tabcolsep}{2pt} 
\begin{tabular}{ccllcllcc}
\hline 
\textbf{Datasets} &
  \multicolumn{3}{c}{\textbf{\begin{tabular}[c]{@{}c@{}}Number of\\ Subjects\end{tabular}}} &
  \multicolumn{3}{c}{\textbf{\begin{tabular}[c]{@{}c@{}}Number of\\ Classes\end{tabular}}} &
  \textbf{\begin{tabular}[c]{@{}c@{}}Class \\ Distribution\end{tabular}} &
  \textbf{\begin{tabular}[c]{@{}c@{}}Subjects'\\ Data\\ Distribution\end{tabular}} \\ \hline
BandX  & \multicolumn{3}{c}{16} & \multicolumn{3}{c}{7}  & Uniform     & Uniform     \\ \hline
WISDM  & \multicolumn{3}{c}{51} & \multicolumn{3}{c}{18} & Uniform     & Uniform     \\ \hline
PAMAP2 & \multicolumn{3}{c}{9}  & \multicolumn{3}{c}{12} & Non-uniform & Non-uniform \\ \hline
\end{tabular}
\vspace{-0.1in}
\end{table}

\noindent\textbf{Pre-processing:} To ensure consistency, we resampled all datasets to $20$ Hz. This involved resampling BandX from $25$ Hz, WISDM from 20 Hz, and PAMAP2 from $100$ Hz. WISDM provided smartwatch IMU sensor data, while PAMAP2's accelerometer and gyroscope readings from wrist sensors have been used. The sliding window technique with 3-second windows and 50\% overlap is applied to capture activity patterns. Each instance was represented as a $(60 x 6)$ matrix, with six channels (three accelerometric, three gyroscopic). This data was fed into the activity classification model.

\vspace{0.05in}
\noindent\textbf{Data Slicing: } Our experiment relies on two distinct dataset types for evaluation of the HAR performance: generalized datasets, comprising data from a wide range of users, and personalized datasets, consisting solely of data from an individual user. To facilitate this, we split the original dataset into two sections. The first section, geared towards generalized analysis, encompasses data from $N-4$ users, where $N$ represents the total number of users in the dataset, and $4$ is a randomly selected number. The second section contains personalized data from four users, specifically designated for evaluating personalized performance metrics.

\subsection{Methodology}
A 1D lightweight CNN model is employed to classify human activities. The model's architecture is depicted in Figure \ref{neuralArchitecture}. Initially, a model is trained using a generalized dataset, serving as the pre-trained model for subsequent personalization. In the personalization phase, a small amount of data from the target user is employed to fine-tune the pre-trained model. Specifically, only the last layer of the pre-trained model undergoes fine-tuning, utilizing a limited amount of data, typically comprising $15$ examples for each activity class. For instance, if the dataset comprises $7$ different activity classes, a total of $105$ examples $(7\:\:classes \times 15\:\:examples$) are utilized for fine-tuning the model.

\subsection{Results and Discussions}
The model is initially trained and evaluated using generalized data in generalized settings. The same model is then assessed in personalized settings, where four personalized datasets are utilized. To mitigate biases, the model's performance is evaluated on each personalized dataset, and the averages are computed. The results are presented in Table \ref{acc_gm}.

\begin{table}[htpb]
    \centering
    \caption{Accuracy of Generalized Model in Generalized (GS) and Personalized (PS) Settings.}
    \label{acc_gm}
    \vspace{-0.1in}
    \setlength{\tabcolsep}{3pt}
    \begin{tabular}{cc|cc|cc}
        \hline
        \multicolumn{2}{c|}{\textbf{BandX }} & \multicolumn{2}{c|}{\textbf{WISDM }} & \multicolumn{2}{c}{\textbf{PAMAP2 }} \\
        \textbf{GS} & \textbf{PS} & \textbf{GS} & \textbf{PS} & \textbf{GS} & \textbf{PS} \\ \hline
        0.9745 & 0.8066 & 0.8339 & 0.7157 & 0.9048 & 0.6007 \\ \hline
    \end{tabular}
     \vspace{-0.1in}
\end{table}

\begin{table}[htpb]
    \centering
    \caption{Accuracy of Generalized Model (GM) and Personalized Model (PM) in Personalized Settings (PS).}
    \label{acc_pm}
    \vspace{-0.1in}
    \setlength{\tabcolsep}{3pt} 
    \begin{tabular}{ll|ll|ll}
        \hline
        \multicolumn{2}{c|}{\textbf{BandX }} & \multicolumn{2}{c|}{\textbf{WISDM }} & \multicolumn{2}{c}{\textbf{PAMAP2 }} \\ 
        \textbf{PS (GM)} & \textbf{PS (PM)} & \textbf{GS (GM)} & \textbf{PS (PM)} & \textbf{GS (GM)} & \textbf{in PS (PM)} \\ \hline
        0.8066 & 0.9659 & 0.7157 & 0.8029 & 0.6007 & 0.8231 \\ \hline
    \end{tabular}
    \vspace{-0.1in}
\end{table}

The analysis reveals that while the generalized model performs well in generalized settings, its accuracy significantly decreases in personalized settings. Specifically, there is a decrease in accuracy of 17.22\% for BandX dataset, 14.17\% for WISDM, and 33.60\% for PAMAP2. These findings underscore the limitations of a generalized HAR model trained on diverse user data. Although such models may perform adequately at a group level, they may lack precision for individual users. Therefore, constructing personalized models tailored to individual users is essential for precise and reliable results.

In contrast, when the model is personalized using fine-tuning with only 15 data samples per class from the generalized model, the performance improves in personalized settings. The accuracy increases by 19.73\% for BandX, 12.18\% for WISDM, and 37\% for PAMAP2, as shown in Table \ref{acc_pm}.

Furthermore, after converting the model to fit the device using TensorFlow Lite, it is found that the compressed model size is less than $15 KB$, and the inference time for a single prediction is less than $200 ms$.

\section{Conclusions}
\label{conc}
This paper presents the development of an on-device personalized HAR system using TinyML, along with methods for automatic deployment of personalized HAR models to target users with minimal calibration via a web application. The HAR method runs continuously on the device, eliminating the need to transfer to other devices for processing ensuring data privacy, low latency, and independence from network connections. By avoiding the need to activate radio modules, a significant energy saving can be achieved, resulting in reduced power consumption and extended battery life, contributing to a more sustainable solution with a lesser carbon footprint. Only a minimal amount of user data is sent to the cloud for model personalization. Users have the option to store inferences in the cloud for further data analysis, if desired. This study aims to enable commercial fitness bands to accurately detect and track a variety of personalized activities directly on the device, while prioritizing user privacy and consent. 

Future plans include enhancing the web application to analyze stored data for spatiotemporal patterns that would provide deeper insights into individual and group activity behaviours.

\bibliographystyle{ACM-Reference-Format}
\bibliography{sample-base}

  \end{document}